%% file: GR-w-GA_PRD.tex
\documentclass[aps,prd,twocolumn,groupedaddress]{revtex4-2}

\usepackage[utf8]{inputenc}
\usepackage[english]{babel}
\usepackage{amsmath,amsfonts,amssymb}
\usepackage{graphicx}
\usepackage{amsthm} 
\usepackage{empheq}
\usepackage{tensor}
\usepackage{hyperref}
\usepackage{subfig} 
\usepackage{xcolor}
\usepackage{cleveref}
\usepackage{bm} 
\usepackage[most]{tcolorbox}
\usepackage{booktabs}
\usepackage{afterpage} 
\usepackage{todonotes} 

\definecolor{graylight}{cmyk}{.30,0,0,.67} 

\numberwithin{equation}{section}

\newcommand{\T}[2]{\tensor{#1}{#2}}
\newcommand{\p}{\partial}
\newcommand{\Tn}[2]{\tensor*{#1}{#2}} 
\newcommand{\g}{\gamma}

\newcommand{\G}{\Gamma}

\newcommand{\dd}{\mathrm{d}}

\crefname{equation}{equation}{equations}
\Crefname{equation}{Equation}{Equations}
\crefrangelabelformat{equation}{(#3#1#4--#5#2#6)}

\crefmultiformat{equation}{equations (#2#1#3}{, #2#1#3)}{#2#1#3}{#2#1#3}
\Crefmultiformat{equation}{Equations (#2#1#3}{, #2#1#3)}{#2#1#3}{#2#1#3}

\begin{document}

\title{General Relativity: New insights from a Geometric Algebra approach}
\author{Pablo Bañón Pérez}
\email[]{pau.banon@physi.uni-heidelberg.de}
\affiliation{Heidelberg University}

\author{Maarten DeKieviet}
\email[]{maarten@physi.uni-heidelberg.de}
\affiliation{Heidelberg University}

\date{\today}

\begin{abstract}
	In this paper, we present a series of techniques to describe General Relativity using Geometric Algebra (GA). We emphasize the physical interpretation of quantities and provide a step-by-step guide for performing calculations. In doing so, we show how GA offers insightful information on the physical meaning of the connection coefficients, the Riemann tensor, and other geometrical quantities.
\end{abstract}

\maketitle

\section{Introduction}

Geometric Algebra (GA) is a powerful language capable of describing a wide variety of fields in physics \cite{Doran2013,Doran1994,Lasenby1996,Lasenby2009,Hestenes2003a_OerstedMedal,Hestenes_A_Unified_Language}. Besides its unifying capacity, GA can simplify the description of phenomena and improve physical insights compared to conventional calculations and differential forms. Previous attempts to describe GR using GA were made by Hestenes \cite{Hestenes_STG_with_GC, Hestenes_Curvature_Calc}, who tried a direct translation of the conventional formalism to GA, and by Doran and Lasenby, who developed a Gauge Theory of Gravity \cite{Doran2013,Hestenes2005,Lasenby1998}. Our work can be considered a complement to these, with a more applied and accessible introduction to the topic.

Some advantages of our description arise from the use of GA, while others stem from the usage of tetrads, hence we call it the tetrad-GA formalism. They include:
\begin{itemize}
	\item \textbf{Notational efficiency}: GA provides a description where the degrees of freedom and symmetries of the objects automatically reflect those of the physical objects they represent.
	\item \textbf{Manifestly covariant, ``coordinate-free" expressions}: GA describes objects by their contracted, ``physical" form instead of their components, establishing a formulation of physical laws that relate physical objects instead of their components.
	\item \textbf{Geometric interpretation of physical objects}: The previous two points result in a formalism that allows easier interpretation of equations compared to tensor calculus or differential forms \cite{GAvsDifferentialForms}.
	\item \textbf{Shortened calculations}: Calculations in the tetrad-GA are often shorter and more transparent than in tensorial calculus or differential forms\cite{GAvsDifferentialForms}.
	\item \textbf{Unification of formalism}: GA can efficiently describe all fields of physics \cite{Doran2013, Lasenby2017}. Working with a single mathematical framework allows for easier integration between fields and generalizations.
	\item \textbf{Decoupling the degrees of freedom}: The use of tetrads separates the degrees of freedom related to the choice of frame from those related to the choice of coordinates, revealing the proper dependencies of objects and facilitating their treatment.
\end{itemize}

It is important to remark that, contrary to the Gauge Theory of Gravity by Doran and Lasenby, which is a theory of gravity in flat space-time, our gravitational theory is conventional General Relativity as developed by Einstein, where gravity is the manifestation of the curvature of space-time, albeit formulated with the tools of GA.

\section{Space-Time Algebra} \label{sec:STA}

A detailed introduction to GA can be found in various sources \cite{Hestenes2003a_OerstedMedal, Macdonald, Macdonald_2002}; we refer the reader to those sources and proceed to present its application to the Minkowski space-time.

Because the notation in GA texts differs from the ones in the GR literature, we wrote an explanation about our notation in \Cref{sec:notation}.

The GA of the Minkowski space-time is called Space-Time Algebra (STA) \cite{Hestenes_STA}. It has the usual metric $\eta_{\mu\nu} = \text{diag}(+,-,-,-)$ and basis vectors $\{\g_\mu\}$ which satisfy the relationship
\begin{equation} \label{eq:Dirac_Algebra}
	\g_\mu \cdot \g_\nu = \eta_{\mu\nu}.
\end{equation}
The basis elements are combined to construct the basis elements of $\mathcal{C}l_{1,3}$ shown in \cref{tab:STA_Basis}. 

\begin{table}[h!]
	\centering
	\begin{tabular}{l | c | c | c | c | c | c }
		Scalars & 1 & & & & & \\ 
		4 vectors & $\gamma_0$ & $\gamma_1$ & $\gamma_2$ & $\gamma_3$ & & \\
		6 bivectors & $\gamma_{10}$ & $\gamma_{20}$ & $\gamma_{30}$ & $\gamma_{23}$ & $\gamma_{31}$ & $\gamma_{12}$ \\
		4 trivectors & $\gamma_{123}$ & $\gamma_{230}$ & $\gamma_{310}$ & $\gamma_{120}$ & & \\ 
		1 Pseudoscalar & $\gamma_{0123}$ & & & & & \\
	\end{tabular}
	\caption{Basis elements of $\mathcal{C}l_{1,3}$, with $\g_{\mu\nu} \equiv \g_\mu \wedge \g_\nu$.}
	\label{tab:STA_Basis}
\end{table}

The reciprocal basis, $\{\g^\mu\}$, is defined by
\begin{equation} \label{eq:STA_Reciprocal_Bas}
	\g_\mu \g^\nu = \Tn{\delta}{^\nu_\mu}.
\end{equation}

\subsection{Rotations and Boosts} \label{sec:rotations-and-boosts}

Due to the mixed signature of the space, in STA we can find two types of basis bivectors: spatial bivectors, with positive square, $(\g_{ij})^2 = -1$, and space-time bivectors, with negative square, $(\g_{0i})^2 = +1$. Where we have introduced the shortened notation $\g_{\mu\nu} \equiv \g_\mu \wedge \g_\nu $.

The set of bivectors of a particular space is a representation of its Lorentz group, where spatial bivectors generate spatial rotations, and space-time bivectors produce boosts \cite{Hestenes2003_STPhysics_With_GA, Hestenes_STA, Dressel2015}. 

To perform a Lorentz transformation of a multivector $M$, we sandwich it between the corresponding rotor $R$
\begin{equation} \label{eq:Rotation}
	M' = RM\tilde{R} = e^{-\frac{\theta}{2}\g_{\mu\nu}} M e^{\frac{\theta}{2} \g_{\nu\mu}},
\end{equation}
where $\theta$ is the parameter of the transformation, and $R$ is a multivector called rotor which can be written as
\begin{equation} \label{eq:Rotor}
	R = e^{-\frac{\theta}{2} \g_{\mu\nu}} = \cos\frac{\theta}{2} - \g_{\mu\nu}\sin\frac{\theta}{2}.,
\end{equation}
and \( \tilde{R} \) denotes the reversion operation, which inverts the order of outer products.

\subsection{Vector Derivative}

The fundamental derivative operator in STA is the \textit{vector derivative} $\nabla$ \cite{Hestenes_GC, Dressel2015, Hestenes_STA}, which can be defined with the reciprocal basis $\g^\mu$ as
\begin{equation} \label{eq:GA_Derivative}
	\nabla = \sum_\mu \g^\nu \p_\mu,
\end{equation}
where $\p_\mu$ is the usual directional derivative $\p_\mu = \p / \p x^\mu$ along the coordinate $x^\mu$.

The vector derivative $\nabla$ has the same form as the Dirac operator constructed with the gamma matrices, but in STA it has the natural interpretation of a vector with derivative operators as components.

An essential feature of $\nabla$ is that it has the algebraic properties of a vector, and we can treat it as such when performing operations\cite{Hitzer2013}.

As an example, we obtain the usual differential operations by application of the inner and outer products over a vector field $v = v(x) = v^\mu(x) \g_\mu$
\begin{itemize}
	\item Geometric product: $\nabla v = \nabla \cdot v + \nabla \wedge v$.
	\item Divergence: $\nabla \cdot v = \p_\mu v^\mu$.
	\item Curl/Exterior derivative: $\nabla \wedge v = \frac{1}{2} \sum_{\mu, \nu=0}^{3} (\p_\mu v^\nu - \p_\nu v^\mu) \g_\mu \wedge \g^\nu$.
	\item Laplacian: $\nabla^2 v = \nabla (\nabla v) = \left(\p^\nu\p_\nu v^\mu \right) \g_\mu$.
\end{itemize}
The associativity properties of the geometric product allow us to construct the Laplacian by making $\nabla$ act on itself, defining a scalar operator which, in contrast with the usual Laplacian, can act over any multivector field.

The possibility of obtaining all differential operators from a single one is unique to Geometric Calculus. It greatly simplifies calculations, improves the geometric interpretation of equations, and unifies algebraic and differential identities \cite{Hestenes_GC, Hitzer2013, Doran2013}.

\section{STA in curved manifolds}

The use of tetrads, defined as orthonormal frames, in curved manifolds combines seamlessly with GA to provide a powerful formalism which we call tetrad-GA.

\begin{figure}
	\centering
	\def\svgwidth{0.4\textwidth}
	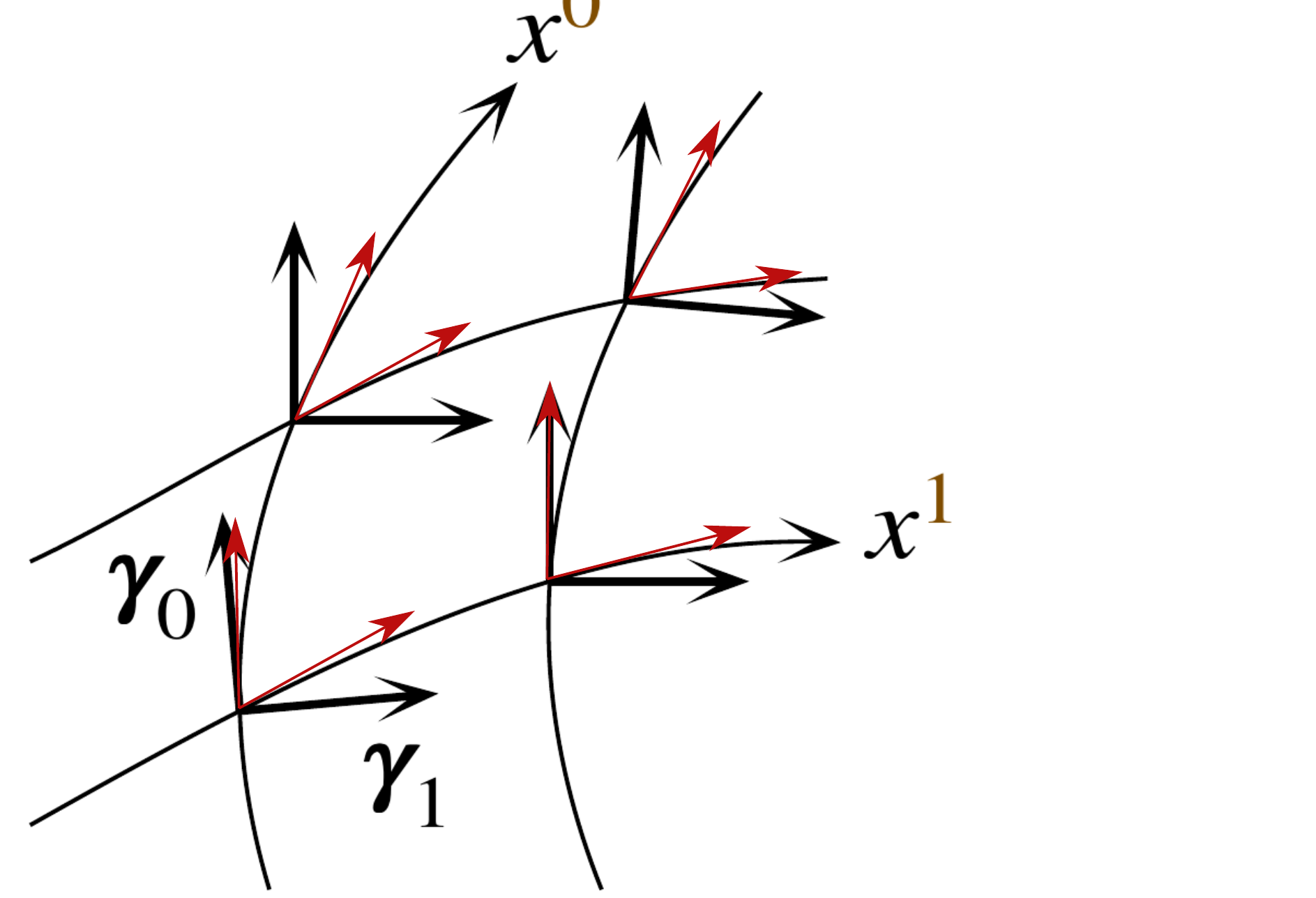
	\caption{Manifold mapped with coordinates $(x^1, x^2)$ and two bases for the tangent space at each point: In red, the coordinate tangent vectors $\{g_\mu\}$, and in black, an orthonormal set of basis vectors $\{\g_m\}$, a tetrad.} 
	\label{fig:Tetrad}
\end{figure}

\subsection{Tetrads with GA}

We denote the coordinate frame of the tangent space \(T_pM\) as \(\{g_\mu\}\). It can be obtained by performing partial derivatives of the coordinate map, \(g_\mu = \p_\mu x\), and it is generally non-orthonormal. This non-orthonormality is encoded in the inner product of the coordinate basis vectors, producing the components of the metric in a certain coordinate choice:
\begin{equation} \label{eq:scalar_prod_metric}
	\T{g}{_\mu} \cdot \T{g}{_\nu} = \T{g}{_{\mu\nu}}.
\end{equation}

We can always perform an orthonormalization of \(\{g_\mu\}\) to obtain an orthonormal frame \(\{\g_m\}\) called a tetrad. A tetrad represents the frame of reference of a local, free-falling observer at point \(p\).

In this paper, we will restrict ourselves to Minkowskian tetrads, whose basis vectors \(\{\g_m\}\) satisfy \cref{eq:Dirac_Algebra}.

The tetrad and coordinate frame are related by a transformation called the vierbein, \(e\indices{^m_\mu}\):
\begin{equation} \label{eq:Definition_vierbein}
	g_\mu = \T{e}{^m_\mu} \g_m.
\end{equation}
The components of the vierbein are determined by the coordinate metric \(g_{\mu\nu}\) and the tetrad metric \(\eta_{mn}\), and can be read directly from the line element using \cref{eq:Definition_vierbein,eq:scalar_prod_metric}:
\begin{equation} \label{eq:Line_element_vierbein}
	\dd s^2 = g_{\mu\nu} \dd x^\mu \dd x^\nu = \eta_{mn} e\indices{^m_\mu} e\indices{^n_\nu} \dd x^\mu \dd x^\nu.
\end{equation}
Effectively, the vierbein can be considered as the positive square root of the metric.

The existence of the reciprocal basis, \(\{g^\mu\}\), related to the coordinate basis \(\{g_\mu\}\) by the musical isomorphism \(g^{\mu\nu} g_\nu = g^\mu\), determines the inverse vierbein \(e\indices{_m^\mu}\), which satisfies:
\begin{equation} \label{eq:Vierbein_Inv_Def}
	\T{e}{^m_\nu} \T{e}{_m^\mu} = \delta^\mu_\nu, \quad \T{e}{^m_\mu} \T{e}{_n^\mu} = \delta^m_n
\end{equation}
and relates the reciprocal coordinate and tetrad bases:
\begin{equation}
	g^\mu = \T{e}{_m^\mu} \g^m.
\end{equation}

The tetrad formalism is enhanced by promoting the local vector space \(T_pM\) to a GA, which we call the geometric tangent space at \(p\), \(GT_pM\), where we can locally construct the same set of \(k\)-vector basis objects as in Minkowski space-time \cref{tab:STA_Basis} and locally apply all the GA techniques \cite{Schindler2019}.

\subsubsection{Covariant derivative and connection bivectors}

The covariant directional derivative of a vector, \(D_\mu a\), comprises the components and the frame's variation. If we decompose the vector in the tetrad frame, \(a = a^m \g_m\), the variation of the frame must correspond to an infinitesimal, proper, orthochronous Lorentz transformation. We can express this transformation as the commutator of the frame with the bivector generator of the transformation, \(\omega_\mu\) (see \Cref{sec:covariant-derivative} for details), to obtain:
\begin{equation} \label{eq:Cov_Dir_Derivative}
	\begin{aligned}
		D_\mu a = \p_\mu a + \frac{1}{2} \left[\omega_\mu, a\right].
	\end{aligned}
\end{equation}
With \(\left[\omega_\mu, a\right] = \omega_\mu a - a \omega_\mu\). \Cref{eq:Cov_Dir_Derivative} is also valid for multivectors \cite{Schindler2019}.

The expression to obtain the connection coefficient bivectors is:
\begin{equation} \label{eq:Connection-coefficients}
	\omega_\mu = \frac{1}{2} \left(g^\lambda \wedge \nabla g_{\mu\lambda} + g_\alpha \wedge \p_\mu g^\alpha\right),
\end{equation}
where \(\nabla = g^\mu \p_\mu\) is the flat space-time vector derivative operator, and the last term is computed as \(\p_\mu g^\alpha = \g^m \p_\mu \T{e}{_m^\alpha}\). We present the derivation of \cref{eq:Connection-coefficients} in \Cref{sec:derivation-connection}. 

In the case of a diagonal metric, and choosing a tetrad frame aligned with the coordinate frame, \cref{eq:Connection-coefficients} reduces to:
\begin{equation} \label{eq:Connection-coefficients-diagonal}
	\omega_\mu = \frac{1}{2} g^\mu \wedge \nabla g_{\mu\mu} = \frac{1}{2} g^\mu \wedge g^\nu \p_\nu g_{\mu\mu}
\end{equation}
without summation over the \(\mu\) indices but over \( \nu \).

Because the connection coefficients map vectors to bivectors, there can only be \(4 \times 6 = 24\) of them, corresponding to 6 possible Lorentz transformations in each of the 4 possible directions of displacement. And if the metric is diagonal and our tetrad is aligned with the coordinate axes, their calculation reduces only to 16 derivatives.

The remaining 16 degrees of freedom needed to cover all 40 of the Christoffel symbols are encoded in the vierbein \(\T{e}{_m^\mu}\), and they are related to changes in norm and relative position of the coordinate basis vectors. A problem not present if our frames are by construction orthonormal.

The bivector coefficients have a clear geometric meaning: \textit{They are the generators of the Lorentz transformation experienced by an inertial frame when parallel transported in a particular direction}.

Compared with the Christoffel formula, \cref{eq:Connection-coefficients,eq:Connection-coefficients-diagonal} are easier to apply, and compared with the \textit{guess-and-check} method of differential forms \cite{Misner1973}, they are systematic and clearer.

If we expand \(\omega_\mu\) in the tetrad basis, we can identify the components of the bivector \(\omega_\mu\) as the spin rotation coefficients:
\begin{equation} \label{eq:Bivector_Rot_coef}
	\omega_\mu = \frac{1}{2} \omega_{mn\mu} \g^m \wedge \g^n.
\end{equation}
And because \(\omega_\mu\) is a bivector, its components are automatically anti-symmetric in their first two indices \(\omega_{mn\mu} = -\omega_{nm\mu}\), which is the expected symmetry for the generator of a Lorentz transformation.

We present the relationship of \(\omega_\mu\) with the Christoffel symbols and Ricci rotation coefficients in \Cref{sec:Ricci-christoffel-coefficients}.

\subsection{The covariant vector derivative operator} \label{sec:covariant-derivatives-ga}

In analogy to the vector derivative in flat space-time \(\nabla\), we can define the covariant derivative operator \cite{Schindler2019}:
\begin{equation}
	D = g^\mu D_\mu.
\end{equation}

Algebraically \(D\) is also a vector, and we can apply the usual algebraic identities to obtain the covariant version of the various differential operations. The most usual ones being:
\begin{itemize}
	\item Covariant directional derivative in the \(a\)-direction: \(\left(b \cdot D\right) a = b^\mu D_\mu a\).
	\item The covariant divergence: \(D \cdot a = D_\mu a^\mu\).
	\item The covariant curl/covariant exterior derivative: \(D \wedge a = \frac{1}{2} \sum_{\mu, \nu=0}^{3} (D_\mu a^\nu - D_\nu a^\mu) g_\mu \wedge g^\nu\).
	\item The covariant Laplacian: \(D^2 a = \left(\sum_{\mu=0}^{3} D_\mu^2\right) a\).
\end{itemize}

These definitions are easily extended to their action over multivectors \cite{Schindler2019}.

\subsection{Riemann Tensor} \label{sec:riemann-tensor}

In a torsion-free space, we can express the commutator of covariant derivatives by the operator \(D \wedge D\):
\begin{equation}
	D \wedge D = g^\nu \wedge g^\mu [D_\nu, D_\mu].
\end{equation}

Applying \(D \wedge D\) to a multivector \(M\) and expanding into components, we obtain the action of the Riemann tensor in GA:
\begin{equation} \label{eq:Cocurl_Multivector}
	\begin{aligned}
		D \wedge D M &= g^\mu \wedge g^\nu [D_\mu, D_\nu] M \\
		&= g^\mu \wedge g^\nu [\mathbf{R}(g_\mu \wedge g_\nu), M],
	\end{aligned}
\end{equation}
where we have defined the Riemann tensor \(\mathbf{R}(g_\mu \wedge g_\nu)\) from the connection coefficients as:
\begin{equation} \label{eq:Riemann_components}
	\mathbf{R}(g_\mu \wedge g_\nu) = \mathbf{R}_{\mu\nu} = \p_\mu \omega_\nu - \p_\nu \omega_\mu + [\omega_\mu, \omega_\nu].
\end{equation}

The Riemann tensor in GA is a map from bivectors to bivectors:
\begin{equation}
	\begin{aligned}
		\mathbf{R}: \Lambda^2(\mathcal{V}) &\rightarrow \Lambda^2(\mathcal{V}), \\
		B \in \Lambda^2(\mathcal{V}) &\mapsto \mathbf{R}(B) \in \Lambda^2(\mathcal{V}).
	\end{aligned}
\end{equation}
We can decompose it in the coordinate frame, to recover its usual components, or into the tetrad frame:
\begin{equation}
	\begin{aligned}
		\mathbf{R}(g_\mu \wedge g_\nu) &= \mathbf{R}_{\mu\nu\alpha\beta} g^\alpha \wedge g^\beta \\
		&= \mathbf{R}_{\mu\nu mn} \g^m \wedge \g^n.
	\end{aligned}
\end{equation}

\begin{figure}
	\centering
	\def\svgwidth{0.4\textwidth}
	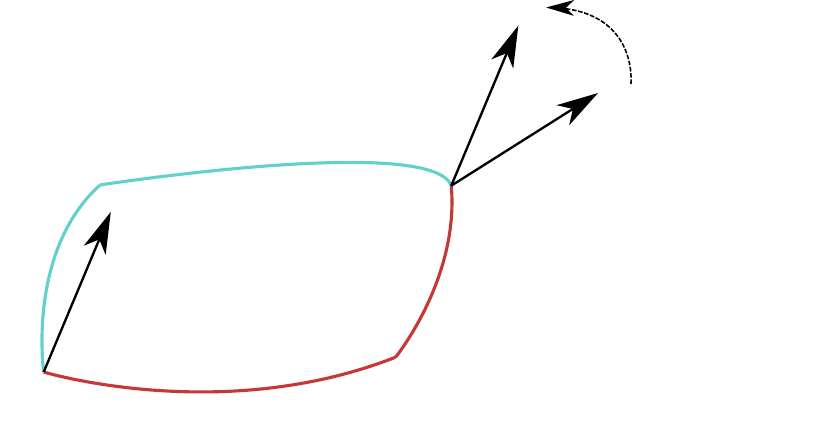
	\caption{Representation of the effect of transporting vector \(v\) to the same point through two different paths. When transported through the red path, \(a \rightarrow b\), the resulting vector is \(v_{ab}\). When transported through the blue path, \(b \rightarrow a\), the resulting vector is \(v_{ba}\). \(v_{ab}\) and \(v_{ba}\) are related by a rotation that is a function of the area spanned between the paths, \(A = a \wedge b\). That function is called the Riemann tensor.} 
	\label{fig:Curvature}
\end{figure}

Considering that bivectors represent areas and are the generators of rotations, the geometrical meaning of the Riemann tensor is apparent: it relates a coordinate differential area, \( g_\mu \wedge g_\nu \) with the rotation experienced by a vector when parallel-transported along its contour, generated by \( \mathbf{R}_{\mu\nu} \), (\cref{fig:Curvature}).

When expressed in the mixed-index form \(\mathbf{R}_{\mu \nu mn}\), the Riemann tensor reveals some of its symmetries straight away:
\begin{enumerate}
	\item The first two components are related to the coordinate area and must be antisymmetric.
	\item The second pair of indices are the basis expansion of the bivector and must also be antisymmetric.
\end{enumerate}
The calculation of the number of degrees of freedom also gets considerably simplified. Because \(\mathbf{R}\) is a function mapping bivectors to bivectors, in a space of 4 dimensions, it can have at most \(6 \times 6 = 36\) degrees of freedom. To reduce them to \( 20 \), we need to consider the protractionless property of the Riemann tensor:
\begin{equation} \label{eq:Riemann_protractionless}
	\p_a \wedge \mathbf{R}(a \wedge b) = 0,
\end{equation}
which comprises a total of \(16\) equations.

\Cref{eq:Riemann_protractionless} implies that \(\mathbf{R}\) is symmetric under the pairwise interchange of indices and that it satisfies the Bianchi algebraic identity:
\begin{equation}
	\mathbf{R}(a \wedge b) \cdot c + \mathbf{R}(c \wedge a) \cdot b + \mathbf{R}(b \wedge c) \cdot a = 0.
\end{equation}
The differential Bianchi identity can be compactly written as:
\begin{equation}
	\dot{D} \wedge \dot{R}(a \wedge b) = g^\mu \left(D_\mu R(a \wedge b) - R( D_\mu (a \wedge b))\right)
\end{equation}
where the overdot to denotes the action of the covariant derivative operator over the Riemann tensor, but not over its arguments \cite{Hestenes_GC}.

This convention can be confusing, but it is necessary to adopt because the vector nature of the derivative operator prevents it from generally commuting with multivectors

\subsection{The Ricci tensor and scalar}
The Ricci tensor appears naturally from \cref{eq:Cocurl_Multivector} when acting on a vector \(a\):
\begin{equation}
	\begin{aligned}
		D \wedge D a &= g^\mu \wedge g^\nu [\mathbf{R}(g_\mu \wedge g_\nu), a] \\
		&= g^\mu \wedge g^\nu \mathbf{R}(g_\mu \wedge g_\nu) \cdot a = R(a),
	\end{aligned}
\end{equation}
where \(R(a) = \T{R}{_{\beta \mu}} a^\beta g^\mu\). In GA, the Ricci tensor is a map from vectors to vectors:
\begin{equation}
	\begin{aligned}
		R: \Lambda^1(\mathcal{V}) &\rightarrow \Lambda^1(\mathcal{V}), \\
		v \in \Lambda^1(\mathcal{V}) &\mapsto R(v) \in \Lambda^1(\mathcal{V}).
	\end{aligned}
\end{equation}
\(R(a)\) is a vector, the dual of which is a 3-volume, \(V = R(a) I\). The Ricci vector quantifies the variation of \(V\) \textit{due to curvature} when displaced in the \(a\)-direction \cite{Loveridge2004}.

The Ricci scalar is obtained by contracting the Ricci vector or directly the Riemann tensor:
\begin{equation}
	\mathcal{R} = g^\mu \cdot R_\mu = (g^\nu \wedge g^\mu) \cdot \mathbf{R}_{\mu\nu}.
\end{equation}

\subsection{Einstein Equations} \label{sec:einstein-equations}

Einstein's tensor in GA is given by
\begin{equation} \label{eq:Einstein_GA}
	G(a) = R(a) - \frac{1}{2}a\mathcal{R},
\end{equation}
and it is also a map from vectors to vectors.

For a given distribution of energy, we can obtain the energy-momentum tensor \( T(a) \) and formulate Einstein's Field Equations in GA as \cite{Hestenes_STG_with_GC}:
\begin{equation} \label{eq:Einstein_Eq_GA}
	G(a) = \kappa T(a) + a \Lambda.
\end{equation}

The energy-momentum tensor in GA, \( T(a) \), is a map from vectors to vectors that returns the 4-momentum that passes through the volume perpendicular to \( a \) \cite{Doran2013}.

The trace-reversed form of \cref{eq:Einstein_Eq_GA} is
\begin{equation}
	R(a) = D \wedge D a = \kappa \left( T(a) - \frac{1}{2} a \text{Tr}(T) \right) + \Lambda a,
\end{equation}
with \(\text{Tr}(T) = \p_a \cdot T(a)\) being the trace of the energy-momentum tensor.

\section{An Illustrative Guide to Calculations} \label{sec:a-guide-to-calculations}

As a step-by-step guide to performing the most common calculations of GR with the tetrad-GA, and to illustrate the geometric content of the objects, we will use the Misner-Thorne metric representing a wormhole through space-time \cite{Wormholes_Morris}.
\begin{equation} \label{eq:Toy_metric}
	\dd s^2 = \dd t^2 - \dd l^2 - (b_0^2 + l^2)(\dd \theta^2 + \sin^2\theta \dd \phi^2),
\end{equation}
where we have set \(c = 1\) and the coordinates have the ranges \(-\infty < t < \infty\), \(-\infty < l < \infty\), \(0 \leq \theta \leq \pi\), \(0 \leq \phi \leq 2\pi\), and \(b_0\) is a constant.

We can directly read the vierbein from the line element, which reduces to its positive “square root” in the case of a diagonal metric:
\begin{equation}
	\T{e}{^m_\mu} = 
	\begin{pmatrix}
		1 & 0 & 0 & 0 \\
		0 & 1 & 0 & 0 \\
		0 & 0 & \sqrt{b_0^2+l^2} & 0 \\
		0 & 0 & 0 & \sqrt{b_0^2+l^2} \sin (\theta ) \\
	\end{pmatrix}.
\end{equation}
We also need its inverse:
\begin{equation}
	\T{e}{_m^\mu} = 
	\begin{pmatrix}
		1 & 0 & 0 & 0 \\
		0 & 1 & 0 & 0 \\
		0 & 0 & \frac{1}{\sqrt{b_0^2+l^2}} & 0 \\
		0 & 0 & 0 & \frac{\csc (\theta )}{\sqrt{b_0^2+l^2}} \\
	\end{pmatrix}.
\end{equation}

The vierbein allows us to calculate the connection coefficient bivectors by using \cref{eq:Connection-coefficients-diagonal}, obtaining:
\begin{equation} \label{eq:Toy_connection_coefficients}
	\begin{aligned}
		\omega_t &= 0 \\
		\omega_l &= 0 \\
		\omega_\theta &= \frac{l}{\sqrt{b_0^2+l^2}} \g_l \wedge \g_\theta \\
		\omega_\phi &= \frac{l \sin (\theta )}{\sqrt{b_0^2+l^2}} \g_l \wedge \g_\phi + \cos (\theta ) \g_\theta \wedge \g_\phi.
	\end{aligned}
\end{equation}

Because \(\omega_t = \omega_l = 0\), an observer parallel-displaced in the \(t\) or \(l\)-directions would experience no rotation or boost. However, an inertial frame parallel transported in the \(\theta\)-direction would rotate in the \(l-\theta\) plane with a rotation speed of \( l/\sqrt{b_0^2+l^2} \) with respect to the local tetrads. Similarly, for the \(\phi\)-direction, where parallel transported frames rotate in the \(l-\phi\) and the \(\theta-\phi\) planes with the coefficients of the bivectors corresponding to the angular velocity of rotation.

We obtain the components of the Riemann tensor \(\mathbf{R}_{\mu\nu}\) with \cref{eq:Riemann_components} and express the result in the tetrad frame for compactness as
\begin{equation}
	\mathbf{R}_{mn} = \frac{b_0^2}{\left(b_0^2+l^2\right)^2} \g_m \wedge \g_n,
\end{equation}
for \(m,n = l, \theta, \phi\) and \(m \neq n\).

Direct contraction with \(\g^m\) easily produces the only non-zero component of the Ricci vectors:
\begin{equation}
	R_{\hat{l}} = \frac{2 b_0^2}{\left(b_0^2+l^2\right)^2} \g_l,
\end{equation}
and further contraction produces the Ricci scalar \(\mathcal{R} = \g^m \cdot R_m\):
\begin{equation}
	\mathcal{R} = \frac{2 b_0^2}{\left(b_0^2+l^2\right)^2}.
\end{equation}

We construct Einstein's vector from \cref{eq:Einstein_GA}:
\begin{equation}
	\begin{aligned}
		G_m &= -\frac{b_0^2}{\left(b_0^2+l^2\right)^2} \g_m, \quad m = \hat{t}, \hat{\theta}, \hat{\phi} \\
		G_{\hat{l}} &= \frac{b_0^2}{\left(b_0^2+l^2\right)^2} \g_l.
	\end{aligned}
\end{equation}

By looking at \cref{eq:Einstein_Eq_GA} with \(\Lambda = 0\), we can obtain the energy-momentum vector necessary to obtain the desired geometry:
\begin{equation}
	\begin{aligned}
		T_m &= -\frac{b_0^2}{8 \pi G \left(b_0^2+l^2\right)^2} \g_m, \quad m = \hat{t}, \hat{\theta}, \hat{\phi} \\
		T_{\hat{l}} &= \frac{b_0^2}{8 \pi G \left(b_0^2+l^2\right)^2} \g_l.
	\end{aligned}
\end{equation}

Because we are working in the tetrad frame, the components of \(T\) are those that the inertial observers would measure. Associating \(T_{\hat{t}}\) with the energy density \(\hat{\rho}(l)\), \(T_{\hat{l}}\) with the radial tension \(\hat{\tau}(l)\), and \(T_{\hat{\theta}}\) and \(T_{\hat{\phi}}\) with the lateral pressure \(\hat{p}_\theta(l)\) and \(\hat{p}_\phi(l)\).

A quick glance at \(T_{\hat{t}}\) shows that \(\hat{\rho}(l) < 0\), which would be a violation of the energy conditions, rendering the energy-momentum necessary to create the desired geometry non-physical.

\section{Conclusions}

In this article, we have reviewed the essential elements of Space-Time Algebra and expanded its application to curved manifolds by generalizing the tangent space from a vector space to a Geometric Algebra (GA).

We have presented the formulation of the usual objects of differential geometry in GA: the connection coefficients, Riemann tensor, Ricci tensor and Ricci scalar, the Einstein tensor, and Einstein's equations.

The GA formulation of GR matches particularly well with the tetrad formalism of GR, significantly simplifying calculations and providing a geometric interpretation of objects that the conventional tensorial treatment lacks.  However, we should note that the commutation properties of objects is not always obvious, which can create a somewhat steep learning curve

The popularization of this knowledge would greatly benefit students due to the simplified calculations and more straightforward physical interpretation, and researchers for its computational power and the simplified connection between fields that GA permits.

\section*{Acknowledgments}

Pablo Bañón Pérez and Maarten DeKieviet would like to thank the Vector-Stiftung, in the framework of the MINT innovation program, and the Heideberg Graduate School For Physics for their financial support.

\bibliography{Biblio}

\vfill
\pagebreak


\appendix

\section{Notation} \label{sec:notation}

We aimed to use common notation with other fields where GA has been applied to make the connection between fields as seamless and intuitive as possible. In this section, we will explain our choice of notation and briefly connect it with other fields.

\begin{itemize}
	\item Our choice of notation for the inner, outer, and geometric product follows the usual convention in GA \cite{Doran2013}.
	\item For the coordinate frame, we chose $\{g_\mu\}$ because their inner product produces the components of the metric, \( g_\mu \cdot g_\nu = g_{\mu\nu} \).
	\item For the tetrad frame, we chose $\{\g_m\}$ because the most common tetrad frame is Minkowskian and it establishes a nice correspondence between Greek and Latin letters to change between coordinate and tetrad quantities. The choice of $\{\g_m\}$ as the basis frame for Minkowski space-time might seem arbitrary until one realizes that the basis vectors of flat space-time can be identified with the Dirac matrices, which allows for a neat interpretation of Dirac theory without complex numbers \cite{Hestenes2003b, Doran2005_Electron_Physics}.
	
	In the GR context, this choice of notation creates a nice correspondence with the treatment of spinors in curved backgrounds, which necessitates tetrads to be properly included.
	
	This notation is also in line with other work treating electromagnetism with GA \cite{Dressel2015} and facilitates the treatment of electromagnetism in curved space-times.
	
	\item Latin middle indices $\{m,n,l,...\}$ refer to tetrad indices, while their Greek counterparts, $\{\mu, \nu, \lambda, ...\}$ refer to coordinate indices. This is in line with the conventional treatment of GR and some of the literature on tetrads.
	
	When an index takes a particular value, we hat the tetrad indices, \( a^{\hat{r}} \), and leave un-hatted the coordinate ones, \( a^r \).
	
	\item The choice of $\omega(g_\mu) = \omega_\mu$ is the usual for connection coefficients in the tetrad formalism and serves two purposes. One, to distinguish them from the Christoffel symbols, denoted by $\G$, and two, to reflect the fact that they are the generators of rotations and their value corresponds to the angular velocity of rotation of a frame displaced in the $g_\mu$ direction.
	
	\item We chose $D = g^\mu D_\mu$ for the covariant vector derivative because in the GA literature where $\nabla$ is used as the vector derivative \textit{in flat space-time}. The reason being that \( \nabla \) can be identified with the Dirac operator and it neatly matches the usual notations for gradient, divergence, and curl: $\nabla\phi$, $\nabla \cdot a$, $\nabla \wedge a$.
	
	\item The use of three different fonts to describe the Riemann tensor, Ricci vector, and Ricci scalar is necessary to differentiate them when their argument is not present.
\end{itemize}


\section{Covariant Derivative} \label{sec:covariant-derivative}

At any given point, the relationship between two tetrad frames is necessarily a transformation of the group \( SO(1,3) \). Since tetrads represent local inertial frames, we can also require that the transformation is orthochronous and respects parity, leaving us with the restricted Lorentz group \( SO^+ (1,3) \).

If we perform a parallel transport of a tetrad frame \(\{\g_m\}\) from the point \(p\) to the point \(q\), the relationship between the local tetrad at \(q\), \(\g_m^{(q)}\), and the transported tetrad \(\g_m^{\prime(q)}\) is necessarily a Lorentz transformation. A Lorentz transformation in GA is performed by sandwiching with the corresponding rotor \(R\):
\begin{equation} \label{eq:Tetrad_transformation}
	\g_m^{\prime(q)} = R \g_m^{(q)} \tilde{R}.
\end{equation}

If the generator of the transformation is the bivector \(\omega_\mu\), and the transformation is infinitesimal with a parameter \(\epsilon\), we can express the rotor \(R\) as
\begin{equation}
	R = \exp\left(\frac{\epsilon}{2} \omega_\mu\right) \approx 1 + \frac{1}{2}\epsilon \omega_\mu,
\end{equation}
and \cref{eq:Tetrad_transformation} reduces to
\begin{equation}
	\g_m^{\prime(q)} = \g_m^{(q)} + \frac{\epsilon}{2}\left[\omega_\mu, \g_m^{(q)}\right],
\end{equation}
where \(\left[\omega_\mu, \g_m^{(q)}\right] = \omega_\mu \g_m^{(q)} - \g_m^{(q)} \omega_\mu\) is called the commutator between multivectors.

Then, we can express the covariant derivative as
\begin{equation}
	\begin{aligned}
		D_\mu a &= (\p_\mu a^m)\g_m + a^m \p_\mu \g_m \\
		&= (\p_\mu a^m)\g_m + a^m \frac{1}{2}\left[\omega_\mu, \g_m\right] \\
		&= \p_\mu a + \frac{1}{2}\left[\omega_\mu, a\right].
	\end{aligned}
\end{equation}


\section{Derivation of Connection Coefficients} \label{sec:derivation-connection}

This derivation was first presented in \cite{snygg_new_2012}. The following is a slight modification of his steps.

We start with a tetrad basis frame expressed in the basis of our coordinate frame:
\begin{equation}
	\g_m = \T{e}{_m^\mu} g_\mu.
\end{equation}
We perform a covariant directional derivative in the \( g_\alpha \) direction:
\begin{equation}
	D_\alpha \g_m = \T{e}{_m^\mu} D_\alpha g_\mu + \left(\p_\alpha \T{e}{_m^\mu}\right) g_\mu.
\end{equation}
where we wrote \(\p_\alpha\) when \(D_\alpha\) acts over a scalar.

We can identify the left-hand side with \( \frac{1}{2} [\omega_\alpha, \g_m] = \omega_\alpha \cdot \g_m \) by definition of \( \omega_\alpha \), \cref{eq:Cov_Dir_Derivative}. And \( D_\alpha g_\mu = \G_{\mu \alpha}^\beta g_\beta \), from the Christoffel symbols definition.

Now we left-multiply by \( \g^m \) to isolate \( \omega_\alpha \):
\begin{equation}
	\g^m \omega_\alpha \cdot \g_m = \g^m \T{e}{_m^\mu} \G_{\mu \alpha}^\beta g_\beta + \g^m \left(\p_\alpha \T{e}{_m^\mu}\right) g_\mu,
\end{equation}
and use the property \( \g^m \left(\g_m \cdot A_r \right) = r A_r \), being \( A_r \) an \( r \)-vector, to simplify the left-hand side. On the right-hand side, we expand the Christoffel symbols into the derivatives of the metric:
\begin{equation}
	\begin{aligned}
		-2 \omega_\alpha &= \T{e}{_m^\mu} \g^m \frac{g^{\beta \lambda}}{2} \left(\p_\mu g_{\alpha \lambda} + \p_\alpha g_{\mu \lambda} - \p_\lambda g_{\alpha \mu}\right) g_\beta \\
		&+ \p_\alpha \left(\T{e}{_m^\mu} \g^m \right) g_\mu.
	\end{aligned}
\end{equation}

Because \( \g^m \) is constant, we pushed it inside the derivative in the last term, and now we can identify the terms \( \T{e}{_m^\mu} \g^m = g^\mu \). Considering that \( \omega_\alpha \) is a bivector, we can discard the terms of grade other than 2 on the right-hand side and write the geometric product as outer products to get:
\begin{equation}
	-2 \omega_\alpha = \frac{1}{2} g^\mu \wedge g^\lambda \left(\p_\mu g_{\alpha \lambda} + \p_\alpha g_{\mu \lambda} - \p_\lambda g_{\alpha \mu}\right) + \left(\p_\alpha g^\mu\right) \wedge g_\mu.
\end{equation}

The \( \p_\alpha g_{\mu \lambda} \) term cancels with \( g^\lambda \wedge g^\mu \) due to symmetry. We can write the remaining term, \( \left(\p_\mu g_{\alpha \lambda} - \p_\lambda g_{\alpha \mu}\right) \), as:
\begin{equation} \label{eq:Appendix_Connection_Coefficients}
	\omega_\alpha = \frac{1}{2} \left( g^\lambda \wedge \nabla g_{\alpha \lambda} + g_\mu \wedge \p_\alpha g^\mu \right).
\end{equation}
This is our final expression to obtain the connection coefficients from the metric and the vierbein, which is hidden in the last term as \( \p_\alpha g^\mu = \g^m \p_\alpha \T{e}{_m^\mu} \), and where \( \nabla = g^\nu \p_\nu \) is the flat space-time vector derivative operator.

In the case of having a diagonal metric, \( g_{\mu \nu} = \text{diag}(g_{\mu \mu}) \), and \( g^{\mu \nu} = \text{diag}((g_{\mu \mu})^{-1}) \), if we choose a tetrad frame aligned with the coordinate frame, then the vierbein is also diagonal with components \( \T{e}{^m_\mu} = \text{diag}((g_{\mu \mu})^{1/2}) \) and its inverse is \( \T{e}{_m^\mu} = \text{diag}((g^{\mu \mu})^{1/2}) = \text{diag}((g_{\mu \mu})^{-1/2}) \). Then, \cref{eq:Appendix_Connection_Coefficients} considerably simplifies:
\begin{equation}
	\begin{aligned}
		g_\mu &= |g_\mu| \g_m \Rightarrow g_\mu \wedge \p_\alpha g^\mu \\
		&= \sum_{\mu} |g_\mu| \left(\p_\alpha |g_\mu|^{-1}\right) \eta^{mm} \g_m \wedge \g_m = 0,
	\end{aligned}
\end{equation}
because \( \g_m \wedge \g_m = 0 \). And, the expression for the connection coefficients reduces to:
\begin{equation} \label{eq:Appendix_Connection_Coefficients-diagonal}
	\omega_\alpha = \frac{1}{2} g^\alpha \wedge \nabla g_{\alpha \alpha} = \frac{1}{2} g^\alpha \wedge g^\mu \p_\mu g_{\alpha \alpha},
\end{equation}
without summation over \(\alpha\).

Applying \cref{eq:Appendix_Connection_Coefficients-diagonal} to a diagonal metric requires a maximum of 16 derivatives to obtain the 4 connection coefficients.


\section{Relation between Spin Rotation Coefficients, Christoffel Symbols, and Ricci Rotation Coefficients} \label{sec:Ricci-christoffel-coefficients}

Because the Christoffel symbols \( \T{\G}{_{\kappa\mu\nu}} \) are not tensors, it is not possible to change their indices using the vierbein \( \T{e}{_m^\mu} \) to relate them with the connection coefficients \( \T{\omega}{_{km\nu}} \),
\begin{equation}
	\G\indices{_{\kappa\mu\nu}} \neq \T{e}{^k_\kappa} \T{e}{^m_\mu} \omega_{km\nu}.
\end{equation}

To obtain their relationship, we start with the definition of the Christoffel symbols and expand in the tetrad frame:
\begin{equation}
	\begin{aligned}
		\G\indices{^\kappa_{\mu\nu}} g_\kappa &= \p_\nu g_\mu = \T{\p}{_\nu} \T{e}{^m_\mu} \g_m \\
		&= (\T{\p}{_\nu} \T{e}{^m_\mu}) \g_m + \T{e}{^m_\mu} \T{\p}{_\nu} \g_m \\
		&= (\T{\p}{_\nu} \T{e}{^l_\mu}) \g_l + \T{e}{^m_\mu} \omega\indices{_m^n_\nu}\g_n \\
		&= (\T{\p}{_\nu} \T{e}{^l_\mu}) \g_l + \T{e}{^m_\mu} \omega\indices{_m^n_\nu} \T{e}{_n^\kappa}g_\kappa \\
		&= \left[\T{e}{_l^\kappa} \T{\p}{_\nu} \T{e}{^l_\mu} + \T{e}{^m_\mu} \T{e}{_n^\kappa} \omega\indices{_m^n_\nu}\right] g_\kappa.
	\end{aligned}
\end{equation}
Thus, we obtain the relationship between the Christoffel symbols and the rotation coefficients:
\begin{equation} \label{eq:Christoffel_Connection_Rel}
	\G\indices{^\kappa_{\mu\nu}} = \T{e}{_l^\kappa} \T{\p}{_\nu} \T{e}{^l_\mu} + \T{e}{^m_\mu} \T{e}{_n^\kappa} \omega\indices{_m^n_\nu}.
\end{equation}

We can invert \cref{eq:Christoffel_Connection_Rel} with the inverse vierbein to obtain the connection coefficients in terms of the Christoffel symbols:
\begin{equation} \label{eq:Connection_Christoffel_Rel}
	\omega\indices{_m^n_\nu} = \T{e}{^n_\kappa} \T{e}{_m^\mu} \G\indices{^\kappa_{\mu\nu}} - \T{e}{^n_\kappa} \T{e}{_m^\mu} \T{e}{_l^\kappa} \T{\p}{_\nu} \T{e}{^l_\mu}.
\end{equation}

The relationship between the spin rotation coefficients \( \omega_{kn\mu} \) and the Ricci connection coefficients \( \omega_{knm} \), corresponding to the covariant directional derivative in the tetrad frame
\begin{equation}
	D_m \g_n = \omega_m \cdot \g_n = (e\indices{_m^\mu} \omega_\mu) \cdot \g_m,
\end{equation}
and it is a direct transformation with the vierbein:
\begin{equation}
	\omega_{kn\mu} = \T{e}{_\mu^m} \omega_{knm}.
\end{equation}

\end{document}

%% file: Tetrad.pdf_tex
\begingroup%
  \makeatletter%
  \providecommand\color[2][]{%
    \errmessage{(Inkscape) Color is used for the text in Inkscape, but the package 'color.sty' is not loaded}%
    \renewcommand\color[2][]{}%
  }%
  \providecommand\transparent[1]{%
    \errmessage{(Inkscape) Transparency is used (non-zero) for the text in Inkscape, but the package 'transparent.sty' is not loaded}%
    \renewcommand\transparent[1]{}%
  }%
  \providecommand\rotatebox[2]{#2}%
  \newcommand*\fsize{\dimexpr\f@size pt\relax}%
  \newcommand*\lineheight[1]{\fontsize{\fsize}{#1\fsize}\selectfont}%
  \ifx\svgwidth\undefined%
    \setlength{\unitlength}{905.52043783bp}%
    \ifx\svgscale\undefined%
      \relax%
    \else%
      \setlength{\unitlength}{\unitlength * \real{\svgscale}}%
    \fi%
  \else%
    \setlength{\unitlength}{\svgwidth}%
  \fi%
  \global\let\svgwidth\undefined%
  \global\let\svgscale\undefined%
  \makeatother%
  \begin{picture}(1,0.68237217)%
    \lineheight{1}%
    \setlength\tabcolsep{0pt}%
    \put(0,0){\includegraphics[width=\unitlength,page=1]{Tetrad.pdf}}%
    \put(0.06273422,0.25313803){\color[rgb]{0,0,0}\makebox(0,0)[lt]{\begin{minipage}{0.09245043\unitlength}\raggedright \end{minipage}}}%
    \put(0.32467709,0.4468437){\color[rgb]{0,0,0}\makebox(0,0)[lt]{\lineheight{1.25}\smash{\begin{tabular}[t]{l}$\textcolor{red}{g_1}$\end{tabular}}}}%
    \put(0,0){\includegraphics[width=\unitlength,page=2]{Tetrad.pdf}}%
    \put(0.24983626,0.52828812){\color[rgb]{0,0,0}\makebox(0,0)[lt]{\lineheight{1.25}\smash{\begin{tabular}[t]{l}$\textcolor{red}{g_0}$\end{tabular}}}}%
    \put(0.14809241,0.4809623){\color[rgb]{0,0,0}\makebox(0,0)[lt]{\lineheight{1.25}\smash{\begin{tabular}[t]{l}$\g_0$\end{tabular}}}}%
    \put(0.26524468,0.32079693){\color[rgb]{0,0,0}\makebox(0,0)[lt]{\lineheight{1.25}\smash{\begin{tabular}[t]{l}$\g_1$\end{tabular}}}}%
    \put(0,0){\includegraphics[width=\unitlength,page=3]{Tetrad.pdf}}%
    \put(0.38631071,0.63394577){\color[rgb]{0,0,0}\makebox(0,0)[lt]{\lineheight{1.25}\smash{\begin{tabular}[t]{l}$x^0$\end{tabular}}}}%
    \put(0.64165002,0.26854645){\color[rgb]{0,0,0}\makebox(0,0)[lt]{\lineheight{1.25}\smash{\begin{tabular}[t]{l}$x^1$\end{tabular}}}}%
  \end{picture}%
\endgroup%

%% file: Curvature_Schema.pdf_tex
\begingroup%
  \makeatletter%
  \providecommand\color[2][]{%
    \errmessage{(Inkscape) Color is used for the text in Inkscape, but the package 'color.sty' is not loaded}%
    \renewcommand\color[2][]{}%
  }%
  \providecommand\transparent[1]{%
    \errmessage{(Inkscape) Transparency is used (non-zero) for the text in Inkscape, but the package 'transparent.sty' is not loaded}%
    \renewcommand\transparent[1]{}%
  }%
  \providecommand\rotatebox[2]{#2}%
  \newcommand*\fsize{\dimexpr\f@size pt\relax}%
  \newcommand*\lineheight[1]{\fontsize{\fsize}{#1\fsize}\selectfont}%
  \ifx\svgwidth\undefined%
    \setlength{\unitlength}{401.3243945bp}%
    \ifx\svgscale\undefined%
      \relax%
    \else%
      \setlength{\unitlength}{\unitlength * \real{\svgscale}}%
    \fi%
  \else%
    \setlength{\unitlength}{\svgwidth}%
  \fi%
  \global\let\svgwidth\undefined%
  \global\let\svgscale\undefined%
  \makeatother%
  \begin{picture}(1,0.50597064)%
    \lineheight{1}%
    \setlength\tabcolsep{0pt}%
    \put(0,0){\includegraphics[width=\unitlength,page=1]{Curvature_Schema.pdf}}%
    \put(0.194325,0.16311401){\color[rgb]{0,0,0}\makebox(0,0)[lt]{\lineheight{1.25}\smash{\begin{tabular}[t]{l}$A=a \wedge b$\end{tabular}}}}%
    \put(0.75726406,0.46411773){\color[rgb]{0,0,0}\makebox(0,0)[lt]{\lineheight{1.25}\smash{\begin{tabular}[t]{l}$R(a \wedge b)$\end{tabular}}}}%
    \put(0.2218504,0.00328911){\color[rgb]{0,0,0}\makebox(0,0)[lt]{\lineheight{1.25}\smash{\begin{tabular}[t]{l}$a$\end{tabular}}}}%
    \put(0.55037951,0.16755362){\color[rgb]{0,0,0}\makebox(0,0)[lt]{\lineheight{1.25}\smash{\begin{tabular}[t]{l}$b$\end{tabular}}}}%
    \put(-0.00101663,0.1613382){\color[rgb]{0,0,0}\makebox(0,0)[lt]{\lineheight{1.25}\smash{\begin{tabular}[t]{l}$b$\end{tabular}}}}%
    \put(0.27512538,0.31808065){\color[rgb]{0,0,0}\makebox(0,0)[lt]{\lineheight{1.25}\smash{\begin{tabular}[t]{l}$a$\end{tabular}}}}%
    \put(0.14282583,0.20129444){\color[rgb]{0,0,0}\makebox(0,0)[lt]{\lineheight{1.25}\smash{\begin{tabular}[t]{l}$v$\end{tabular}}}}%
    \put(0.50523989,0.41398062){\color[rgb]{0,0,0}\makebox(0,0)[lt]{\lineheight{1.25}\smash{\begin{tabular}[t]{l}$v_{ba}$\end{tabular}}}}%
    \put(0.630292,0.30873237){\color[rgb]{0,0,0}\makebox(0,0)[lt]{\lineheight{1.25}\smash{\begin{tabular}[t]{l}$v_{ab}$\end{tabular}}}}%
    \put(0.70310116,0.32382691){\color[rgb]{0,0,0}\makebox(0,0)[lt]{\begin{minipage}{0.17225582\unitlength}\raggedright \end{minipage}}}%
  \end{picture}%
\endgroup%